\documentclass[runningheads]{llncs}

\usepackage{url}
\usepackage{graphicx}
\usepackage{listings}
\usepackage{todonotes}
\usepackage{booktabs}
\usepackage{xcolor}
\usepackage{hyperref}
\usepackage{graphicx}
\usepackage{subcaption}
\usepackage{upquote}
\usepackage{orcidlink}

\begin{document}

\title{Pipeline Inspection, Visualization, \\ and Interoperability in PyTerrier}
\titlerunning{Pipeline Inspection, Visualization, and Interoperability}
\author{Emmanouil Georgios Lionis\orcidlink{0009-0004-3931-9657} \and Craig Macdonald\orcidlink{0000-0003-3143-279X} \and Sean MacAvaney\orcidlink{0000-0002-8914-2659}}
\institute{University of Glasgow, UK \\
\email{e.lionis.1@research.gla.ac.uk\\ \{Craig.Macdonald,Sean.MacAvaney\}@glasgow.ac.uk}}
\maketitle

\begin{abstract}
PyTerrier provides a declarative framework for building and experimenting with Information Retrieval (IR) pipelines. In this demonstration, we highlight several recent pipeline operations that improve their ability to be programmatically inspected, visualized, and integrated with other tools (via the Model Context Protocol, MCP). These capabilities aim to make it easier for researchers, students, and AI agents to understand and use a wide array of IR pipelines.
\end{abstract}

\section{Introduction}
\begin{figure}
\centering
{\scriptsize
\begin{verbatim}
index = pt.Artifact.from_hf('pyterrier/msmarco-passage.terrier')
pipeline = pta.RRFusion(
    index.bm25(),
    SequentialDependence() >> index.bm25(),
) >> index.text_loader() >> MonoT5() >> T5FiD('terrierteam/t5fid_base_nq')
\end{verbatim}
}
\vspace{-0.75em}
(a) Python code to declare the pipeline (imports omitted for brevity).
\vspace{0.5em}

\includegraphics[width=\linewidth]{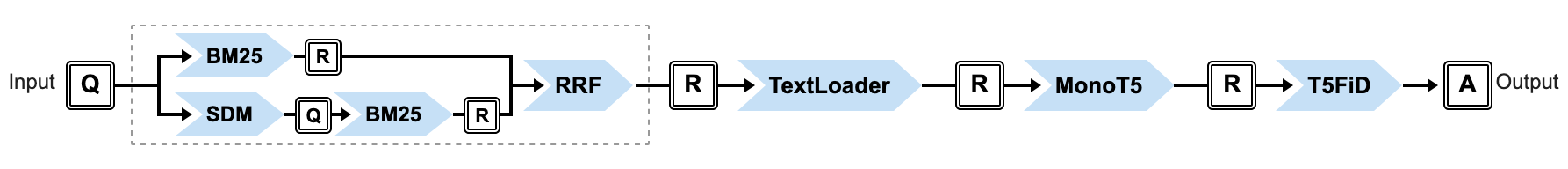}

\vspace{-0.5em}
(b) Generated schematic for the pipeline defined in (a).
\vspace{-0.5em}
\caption{Example PyTerrier pipeline that performs retrieval, fusion, re-ranking, and answer generation stages. The pipeline is shown both as (a) declarative Python code and (b) the corresponding generated schematic.}
\label{fig:pipeline}
\end{figure}

PyTerrier~\cite{DBLP:conf/ictir/MacdonaldT20} provides an interface between state-of-the-art retrieval components, including highly-efficient inverted indexes~\cite{DBLP:conf/sigir/MacAvaneyM22}, generative re-rankers~\cite{DBLP:journals/corr/abs-2412-05339}, and retrieval-augmented generation methods~\cite{DBLP:conf/sigir/MacdonaldFPM25}. Users can combine these components into pipelines (Section~\ref{sec:background}) through a declarative Python syntax. For example, Figure~\ref{fig:pipeline}(a) shows a sophisticated pipeline that performs a variety of operations, including lexical retrieval, query reformulation, result fusion, neural re-ranking, and answer generation. 

This demonstration presents three recent enhancements to PyTerrier pipelines designed to make them easier to understand and interact with. First, we introduce a core \textit{inspection} mechanism (Section~\ref{sec:inspect}), which provides programmatic access to information about pipelines, including input/output specifications, constituent components, and other settings. Building on this foundation, we provide interactive pipeline visualizations that we call \textit{schematics} (Section~\ref{sec:schematics}, Figure~\ref{fig:pipeline}(b)). Schematics are automatically rendered for pipelines in Python notebooks to help users understand the pipelines they construct. They are also integrated into the documentation to serve illustrative examples and can be rendered \textit{ad hoc} for other purposes. Finally, we enable users to expose PyTerrier pipelines as tools via a Model Context Protocol (MCP) server~\cite{misc:mcp2024}, allowing other systems to interact with them through MCP's unified API (Section~\ref{sec:mcp}). For instance, the MCP feature enables pipelines to be easily integrated as components of agentic systems for research and deployment, or used as tools through systems such as Copilot\footnote{\scalebox{0.8}{\url{https://github.com/features/copilot}}} or Cursor\footnote{\scalebox{0.8}{\url{https://cursor.com/}}}.

\section{Background: PyTerrier Pipelines}
\label{sec:background}
\vspace{-1em}

The core components of pipelines in PyTerrier are called transformers. A transformer provides a function that modifies the current \textit{state} of one or more search requests. For example, a BM25 transformer accepts search queries and returns the search results from an index. The state is represented as a relation (i.e., a data table such as a Pandas DataFrame or a list of Python dictionaries). PyTerrier's data model defines a variety of typical column definitions that represent different stages of an engine's processing as shown in Table~\ref{tab:types}. These configurations can be extended with additional information as needed by transformers (e.g., a document's text or a query's vector). PyTerrier hosts components for a wide variety of IR pipeline operations, including lexical retrieval, dense retrieval, learned sparse retrieval, query reformulation, and answer generation.

\begin{table}[b]
\centering
\caption{PyTerrier data model columns for typical stages of a search engine.}
\label{tab:types}
\scalebox{0.7}{
\begin{tabular}{lcl}
\toprule
\bf Name & \bf Abbr. & \bf Columns \\
\midrule
Query Frame & Q & \texttt{qid} (PK, query identifier), \texttt{query} \\
Document Frame & D & \texttt{docno} (PK, document identifier), \texttt{text} \\
Result Frame & R & \texttt{qid} (PK), \texttt{docno} (PK), \texttt{score}, \texttt{rank}\\
Answer Frame & A & \texttt{qid} (PK), \texttt{qanswer} \\
\bottomrule
\end{tabular}
}
\end{table}

PyTerrier's data model enables transformers to be combined together into larger pipelines. In a typical sequential pipeline (represented as \texttt{A >> B}), the output of one transformer (\texttt{A}) can be fed as input to the next transformer (\texttt{B}). Other types of pipelines include linear score combination (e.g., \texttt{A + B}) and reciprocal rank fusion~\cite{DBLP:conf/sigir/CormackCB09} (e.g., \texttt{RRFusion(A, B)}). In all cases, it is important that the pipelines are combined in a way where the input/output specifications are compatible with one another. For instance, if a pipeline is constructed that does not provide the text of a document to a neural re-ranking method that relies on it, the pipeline will raise an error when executed. This demonstration highlights new features that overcome shortcomings of these pipelines, namely the ability to programmatically inspect and validate the inputs/outputs of transformers, the ability to visualize them, and the ability to use them as tools for an LLM.

\section{Inspecting Pipelines}\label{sec:inspect}
\vspace{-1em}

PyTerrier transformers can be defined with as little as a function that maps one relation to another. This flexibility makes implementing transformers simple and allows arbitrary pipelines to be constructed, but can result in incompatible pipelines. Without knowing the input and output specifications of individual transformers, it can be challenging to know whether a pipeline is compatible. Previously the only way to check these would be to check the documentation or run an example through and check the outputs (or interpret the resulting error).

To overcome this inconvenience, we introduce a new \texttt{pt.inspect}
module that allows for the programmatic inspection of transformers and pipelines. The module provides functions that return which columns are required as inputs (e.g., a BM25 retriever expects \texttt{qid} and \texttt{query}, while a FAISS~\cite{DBLP:journals/corr/abs-2401-08281]} dense retriever expects \texttt{qid} and \texttt{query\_vec}) and which columns are returned for the provided inputs (e.g., BM25 returns \texttt{qid}, \texttt{query}, \texttt{docno}, \texttt{rank}, and \texttt{score}). Other functions in the module provide additional information about transformers, such as its attributes and any other transformers that it invokes (subtransformers). In most cases, \texttt{pt.inspect} identifies these automatically\footnote{\scalebox{0.8}{The details are too intricate to describe here, but are covered in the \href{https://pyterrier.readthedocs.io/en/latest/inspect.html}{documentation}.}} by leveraging best practices in transformer implementation (e.g., performing input validation). To provide additional clarity on the implementation, a notebook is included in PyTerrier that demonstrates the inspection functionality.\footnote{\label{fn:inspect}\scalebox{0.8}{\url{https://github.com/terrier-org/pyterrier/blob/master/examples/notebooks/inspect-demo.ipynb}}}

Transformer inspection has a number of applications. It enables pipeline {\em validation} to ensure that a pipeline is fully compatible (which is now applied automatically before experiments are conducted). Inspection also helps in visualization and interoperability, which are covered in the following two sections.

\section{Visualizing Pipelines with Schematics}
\label{sec:schematics}
\vspace{-1em}

As shown in Figure~\ref{fig:pipeline}(a), pipeline definitions can get complex, especially when they involve multiple components and stages. Therefore, we added a new pipeline visualization feature that renders pipelines as interactive HTML snippets. We call these visitations schematics. Schematics render the pipeline as a sequence of transformers and the data between then, allowing users to better understand the transformers in a pipeline and how data flows between them. Hovering over elements provides additional information, like specific information about columns or transformer settings. Schematics leverage the inspection features covered in Section~\ref{sec:inspect}. A non-interactive screenshot of a schematic is shown in Figure~\ref{fig:pipeline}(b).

Although schematics can be displayed anywhere that can render HTML, we have applied them in a couple of convenient places. First, they are automatically rendered in Python notebooks when the output of a cell is a pipeline, aiding in the construction of pipelines in interactive sessions. We also render them in several places throughout PyTerrier's documentation to better demonstrate core concepts of the platform. PyTerrier contains an interactive notebook containing schematics for a range of pipelines.\footnote{\label{fn:schematic}\scalebox{0.8}{\url{https://github.com/terrier-org/pyterrier/blob/master/examples/notebooks/schematics-demo.ipynb}}}

\section{Pipeline Interoperation with MCP}
\label{sec:mcp}
\vspace{-1em}

\begin{figure}[t]
    \centering
    \includegraphics[width=0.75\linewidth]{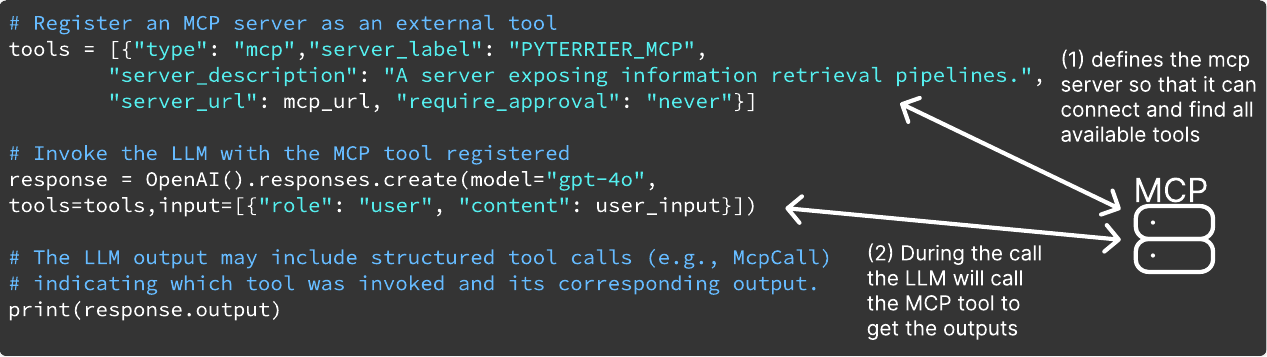}
    \vspace{-1em}
    \caption{Example usage of the MCP server through the  OpenAI API.}
    \label{fig:mcp_tool}
\end{figure}

Another limitation of PyTerrier pipelines is the need for a Python runtime to execute them. Increasingly, external ``tools'' are invoked directly by Large Language Models (LLMs)~\cite{DBLP:conf/nips/SchickDDRLHZCS23}. The open-source Model Context Protocol (MCP)~\cite{misc:mcp2024} is a popular protocol that standardizes LLM access to these tools by exposing them via HTTP endpoints. Despite MCP's original design for LLMs, its scope is much broader and has been referred to as a ``universal plugin ecosystem'' because it allows for simplified interoperation between virtually any Internet-connected system~\cite{misc:plugin2025}. This quality makes MCP a natural choice to improve PyTerrier's interoperability with systems outside of Python.

To this end, we built a new package that exposes PyTerrier pipelines via MCP, thereby allowing a growing ecosystem of MCP-compatible tools to use them. The MCP server exposes HTTP endpoints for PyTerrier pipelines along with their associated metadata, such as natural-language descriptions and input/output specifications (using the new inspection functionality described in Section~\ref{sec:inspect}).
The code for the MCP server is publicly available\footnote{\scalebox{0.8}{\url{https://github.com/terrierteam/pyterrier-server}}}.

The MCP server is useful in a variety of settings. First, it can be used programmatically by researchers or practitioners as components in agentic systems. For instance, one can enable pipeline access for LLMs using the OpenAI\footnote{\scalebox{0.8}{\url{https://platform.openai.com/docs/overview}}} package, as shown in Figure~\ref{fig:mcp_tool}. Alternatively, MCP-exposed PyTerrier pipelines can be leveraged by users directly through systems like the Copilot extension in Visual Studio Code. This allows users to leverage PyTerrier pipelines in their day-to-day activities.

\begin{figure}[t]  
    \centering
\includegraphics[width=\linewidth]{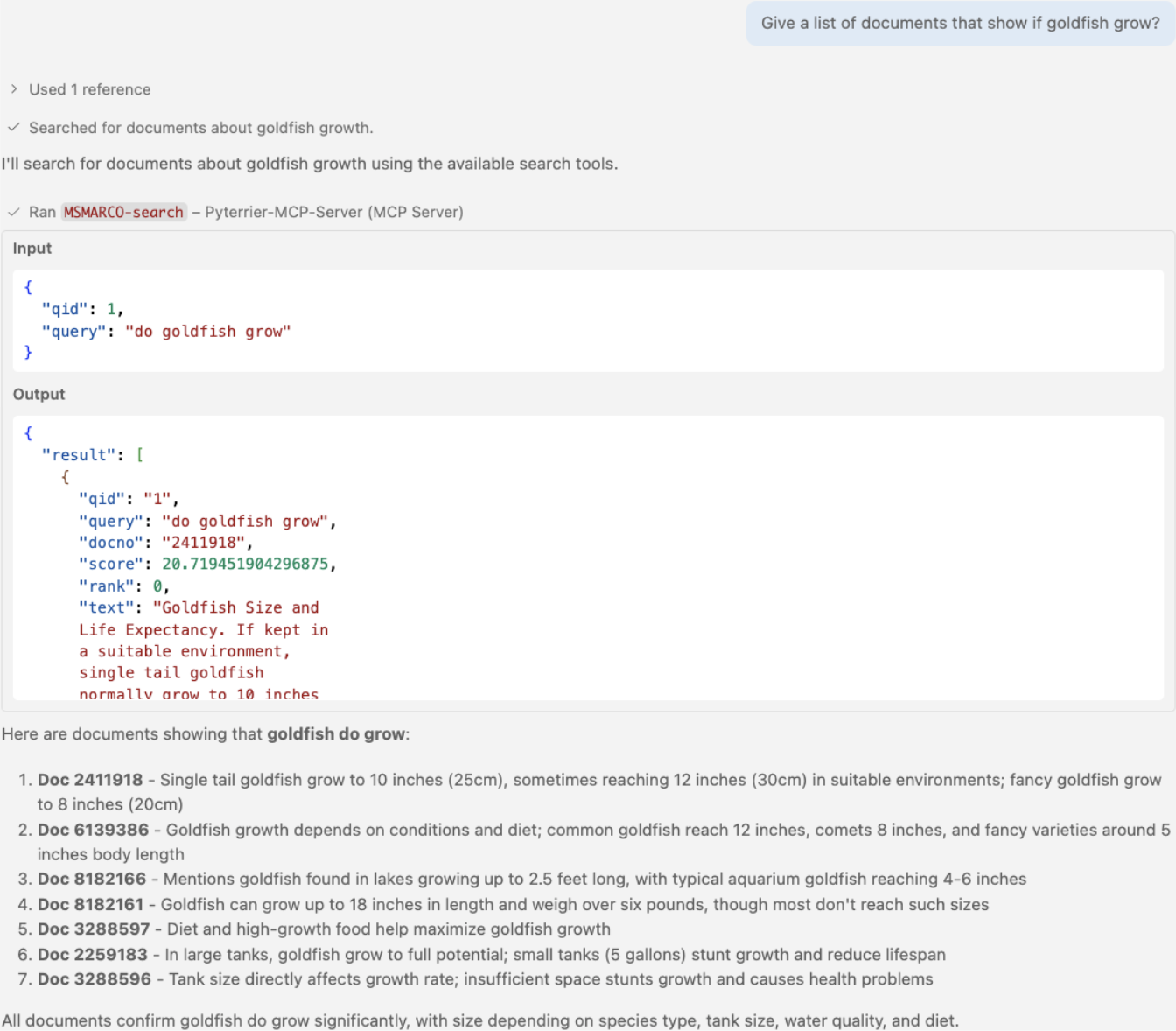}
    \caption{A PyTerrier MCP server interaction example through Copilot.}
    \label{fig:search-mcp}
\end{figure}

To demonstrate the flexibility of the MCP server across different retrieval and reasoning paradigms, we showcase its use with three representative pipelines: (i) a traditional BM25~\cite{DBLP:journals/ftir/RobertsonZ09} retrieval pipeline implemented using PISA~\cite{DBLP:conf/sigir/MalliaSMS19,DBLP:conf/sigir/MacAvaneyM22}; (ii) a retrieval-augmented generation (RAG) system employing a lightweight Fusion-in-Decoder model for direct question answering~\cite{DBLP:conf/eacl/IzacardG21,DBLP:conf/nips/LewisPPPKGKLYR020,DBLP:conf/sigir/MacdonaldFPM25}; and (iii) a Doc2Query-based model designed to support question decomposition~\cite{DBLP:conf/ecir/GospodinovMM23,macdonald2020declarative}. Given these pipelines, we provide a demo video\footnote{\label{fn:mcp}\scalebox{0.8}{\url{https://github.com/terrierteam/pyterrier-server/blob/main/videos/demo.gif}}} illustrating the programmatic implementation via a web-based interface, as well as a demonstration of day-to-day usage through Copilot (Figure~\ref{fig:search-mcp}).

\section{Conclusions}
\label{sec:conclusion}
\vspace{-1em}

This demonstration covers new features in PyTerrier that improve the inspection, visualization, and interoperability of pipelines. ECIR attendees will be able to interact with each of these features through notebooks and a web interface in the demonstration.

\section*{Disclosure of Interests}
The authors have no competing interests to declare that are relevant to the content of this article.

\bibliographystyle{splncs04}
\bibliography{references}

@inproceedings{macdonald2020declarative,
  title={Declarative experimentation in information retrieval using PyTerrier},
  author={Macdonald, Craig and Tonellotto, Nicola},
  booktitle={Proceedings of the 2020 ACM SIGIR on International Conference on Theory of Information Retrieval},
  pages={161--168},
  year={2020}
}

@inproceedings{DBLP:conf/ecir/GospodinovMM23,
  author       = {Mitko Gospodinov and
                  Sean MacAvaney and
                  Craig Macdonald},
  editor       = {Jaap Kamps and
                  Lorraine Goeuriot and
                  Fabio Crestani and
                  Maria Maistro and
                  Hideo Joho and
                  Brian Davis and
                  Cathal Gurrin and
                  Udo Kruschwitz and
                  Annalina Caputo},
  title        = {Doc2Query-: When Less is More},
  booktitle    = {Advances in Information Retrieval - 45th European Conference on Information
                  Retrieval, {ECIR} 2023, Dublin, Ireland, April 2-6, 2023, Proceedings,
                  Part {II}},
  series       = {Lecture Notes in Computer Science},
  volume       = {13981},
  pages        = {414--422},
  publisher    = {Springer},
  year         = {2023},
  url          = {https://doi.org/10.1007/978-3-031-28238-6\_31},
  doi          = {10.1007/978-3-031-28238-6\_31},
  bibsource    = {dblp computer science bibliography, https://dblp.org}
}

@inproceedings{DBLP:conf/nips/LewisPPPKGKLYR020,
  author       = {Patrick Lewis and
                  Ethan Perez and
                  Aleksandra Piktus and
                  Fabio Petroni and
                  Vladimir Karpukhin and
                  Naman Goyal and
                  Heinrich K{\"{u}}ttler and
                  Mike Lewis and
                  Wen{-}tau Yih and
                  Tim Rockt{\"{a}}schel and
                  Sebastian Riedel and
                  Douwe Kiela},
  editor       = {Hugo Larochelle and
                  Marc'Aurelio Ranzato and
                  Raia Hadsell and
                  Maria{-}Florina Balcan and
                  Hsuan{-}Tien Lin},
  title        = {Retrieval-Augmented Generation for Knowledge-Intensive {NLP} Tasks},
  booktitle    = {Advances in Neural Information Processing Systems 33: Annual Conference
                  on Neural Information Processing Systems 2020, NeurIPS 2020, December
                  6-12, 2020, virtual},
  year         = {2020},
  url          = {https://proceedings.neurips.cc/paper/2020/hash/6b493230205f780e1bc26945df7481e5-Abstract.html},
  bibsource    = {dblp computer science bibliography, https://dblp.org}
}

@inproceedings{DBLP:conf/eacl/IzacardG21,
  author       = {Gautier Izacard and
                  Edouard Grave},
  editor       = {Paola Merlo and
                  J{\"{o}}rg Tiedemann and
                  Reut Tsarfaty},
  title        = {Leveraging Passage Retrieval with Generative Models for Open Domain
                  Question Answering},
  booktitle    = {Proceedings of the 16th Conference of the European Chapter of the
                  Association for Computational Linguistics: Main Volume, {EACL} 2021,
                  Online, April 19 - 23, 2021},
  pages        = {874--880},
  publisher    = {Association for Computational Linguistics},
  year         = {2021},
  url          = {https://doi.org/10.18653/v1/2021.eacl-main.74},
  doi          = {10.18653/V1/2021.EACL-MAIN.74},
  bibsource    = {dblp computer science bibliography, https://dblp.org}
}

@inproceedings{DBLP:conf/sigir/MacdonaldFPM25,
  author       = {Craig Macdonald and
                  Jinyuan Fang and
                  Andrew Parry and
                  Zaiqiao Meng},
  editor       = {Nicola Ferro and
                  Maria Maistro and
                  Gabriella Pasi and
                  Omar Alonso and
                  Andrew Trotman and
                  Suzan Verberne},
  title        = {Constructing and Evaluating Declarative {RAG} Pipelines in PyTerrier},
  booktitle    = {Proceedings of the 48th International {ACM} {SIGIR} Conference on
                  Research and Development in Information Retrieval, {SIGIR} 2025, Padua,
                  Italy, July 13-18, 2025},
  pages        = {4035--4040},
  publisher    = {{ACM}},
  year         = {2025},
  url          = {https://doi.org/10.1145/3726302.3730150},
  doi          = {10.1145/3726302.3730150},
  bibsource    = {dblp computer science bibliography, https://dblp.org}
}

@inproceedings{DBLP:conf/sigir/MalliaSMS19,
  author       = {Antonio Mallia and
                  Michal Siedlaczek and
                  Joel M. Mackenzie and
                  Torsten Suel},
  editor       = {Ryan Clancy and
                  Nicola Ferro and
                  Claudia Hauff and
                  Jimmy Lin and
                  Tetsuya Sakai and
                  Ze Zhong Wu},
  title        = {{PISA:} Performant Indexes and Search for Academia},
  booktitle    = {Proceedings of the Open-Source {IR} Replicability Challenge co-located
                  with 42nd International {ACM} {SIGIR} Conference on Research and Development
                  in Information Retrieval, OSIRRC@SIGIR 2019, Paris, France, July 25,
                  2019},
  series       = {{CEUR} Workshop Proceedings},
  volume       = {2409},
  pages        = {50--56},
  publisher    = {CEUR-WS.org},
  year         = {2019},
  url          = {https://ceur-ws.org/Vol-2409/docker08.pdf},
  bibsource    = {dblp computer science bibliography, https://dblp.org}
}

@article{DBLP:journals/ftir/RobertsonZ09,
  author       = {Stephen E. Robertson and
                  Hugo Zaragoza},
  title        = {The Probabilistic Relevance Framework: {BM25} and Beyond},
  journal      = {Found. Trends Inf. Retr.},
  volume       = {3},
  number       = {4},
  pages        = {333--389},
  year         = {2009},
  url          = {https://doi.org/10.1561/1500000019},
  doi          = {10.1561/1500000019},
  bibsource    = {dblp computer science bibliography, https://dblp.org}
}

@inproceedings{DBLP:conf/ictir/MacdonaldT20,
  author       = {Craig Macdonald and
                  Nicola Tonellotto},
  editor       = {Krisztian Balog and
                  Vinay Setty and
                  Christina Lioma and
                  Yiqun Liu and
                  Min Zhang and
                  Klaus Berberich},
  title        = {Declarative Experimentation in Information Retrieval using PyTerrier},
  booktitle    = {{ICTIR} '20: The 2020 {ACM} {SIGIR} International Conference on the
                  Theory of Information Retrieval, Virtual Event, Norway, September
                  14-17, 2020},
  pages        = {161--168},
  publisher    = {{ACM}},
  year         = {2020},
  url          = {https://doi.org/10.1145/3409256.3409829},
  doi          = {10.1145/3409256.3409829},
  bibsource    = {dblp computer science bibliography, https://dblp.org}
}

@inproceedings{DBLP:conf/nips/SchickDDRLHZCS23,
  author       = {Timo Schick and
                  Jane Dwivedi{-}Yu and
                  Roberto Dess{\`{\i}} and
                  Roberta Raileanu and
                  Maria Lomeli and
                  Eric Hambro and
                  Luke Zettlemoyer and
                  Nicola Cancedda and
                  Thomas Scialom},
  editor       = {Alice Oh and
                  Tristan Naumann and
                  Amir Globerson and
                  Kate Saenko and
                  Moritz Hardt and
                  Sergey Levine},
  title        = {Toolformer: Language Models Can Teach Themselves to Use Tools},
  booktitle    = {Advances in Neural Information Processing Systems 36: Annual Conference
                  on Neural Information Processing Systems 2023, NeurIPS 2023, New Orleans,
                  LA, USA, December 10 - 16, 2023},
  year         = {2023},
  url          = {http://papers.nips.cc/paper\_files/paper/2023/hash/d842425e4bf79ba039352da0f658a906-Abstract-Conference.html},
  bibsource    = {dblp computer science bibliography, https://dblp.org}
}

@misc{misc:mcp2024,
  title = {Introducing the Model Context Protocol},
  howpublished = {\url{https://www.anthropic.com/news/model-context-protocol}},
  author={Anthropic},
  year={2024}
}

@misc{misc:plugin2025,
  title={{MCP}: An (Accidentally) Universal Plugin System},
  author={Scott Werner},
  howpublished = {\url{https://worksonmymachine.ai/p/mcp-an-accidentally-universal-plugin}},
  year={2025}
}

@inproceedings{DBLP:conf/sigir/MacAvaneyM22,
  author       = {Sean MacAvaney and
                  Craig Macdonald},
  editor       = {Enrique Amig{\'{o}} and
                  Pablo Castells and
                  Julio Gonzalo and
                  Ben Carterette and
                  J. Shane Culpepper and
                  Gabriella Kazai},
  title        = {A Python Interface to PISA!},
  booktitle    = {{SIGIR} '22: The 45th International {ACM} {SIGIR} Conference on Research
                  and Development in Information Retrieval, Madrid, Spain, July 11 -
                  15, 2022},
  pages        = {3339--3344},
  publisher    = {{ACM}},
  year         = {2022},
  url          = {https://doi.org/10.1145/3477495.3531656},
  doi          = {10.1145/3477495.3531656},
  bibsource    = {dblp computer science bibliography, https://dblp.org}
}

@inproceedings{DBLP:conf/sigir/CormackCB09,
  author       = {Gordon V. Cormack and
                  Charles L. A. Clarke and
                  Stefan B{\"{u}}ttcher},
  editor       = {James Allan and
                  Javed A. Aslam and
                  Mark Sanderson and
                  ChengXiang Zhai and
                  Justin Zobel},
  title        = {Reciprocal rank fusion outperforms condorcet and individual rank learning
                  methods},
  booktitle    = {Proceedings of the 32nd Annual International {ACM} {SIGIR} Conference
                  on Research and Development in Information Retrieval, {SIGIR} 2009,
                  Boston, MA, USA, July 19-23, 2009},
  pages        = {758--759},
  publisher    = {{ACM}},
  year         = {2009},
  url          = {https://doi.org/10.1145/1571941.1572114},
  doi          = {10.1145/1571941.1572114},
  bibsource    = {dblp computer science bibliography, https://dblp.org}
}

@article{DBLP:journals/corr/abs-2412-05339,
  author       = {Kaustubh D. Dhole},
  title        = {PyTerrier-GenRank: The PyTerrier Plugin for Reranking with Large Language
                  Models},
  journal      = {CoRR},
  volume       = {abs/2412.05339},
  year         = {2024},
  url          = {https://doi.org/10.48550/arXiv.2412.05339},
  doi          = {10.48550/ARXIV.2412.05339},
  eprinttype    = {arXiv},
  eprint       = {2412.05339},
  bibsource    = {dblp computer science bibliography, https://dblp.org}
}

\end{document}